# A Coherent Integration Method Based on Radon Non-uniform FRFT for Random Pulse Repetition Interval (RPRI) Radar

Tian Jing, Xia Xiang-Gen, *Fellow, IEEE*, Yang Gang, Cui Wei, and Wu Si-Liang

*Abstract*—To solve the range cell migration (RCM) and spectrum spread during the integration time induced by the motion of a target, this paper proposes a new coherent integration method based on Radon non-uniform FRFT (NUFRFT) for random pulse repetition interval (RPRI) radar. In this method, RCM is eliminated via searching in the motion parameters space and the spectrum spread is resolved by using NUFRFT. Comparisons with other popular methods, moving target detection (MTD), Radon-Fourier transform (RFT), and Radon-Fractional Fourier Transform (RFRFT) are performed. The simulation results demonstrate that the proposed method can detect the moving target even in low SNR scenario and is superior to the other two methods.

## I. INTRODUCTION

IN modern warfare, with the development of electromagnetic interference technique, how to improve jamming suppression ability has attracted considerable attention in the past decades. Fortunately, waveform design [1-4] can be used to achieve this goal, such as random pulse repetition interval (RPRI) signals. However, new problems, such as range cell migration (RCM) [5] and spectrum spread [6], are introduced during long-time coherent integration. We have to resolve these problems to realize coherent integration and further improve the detection performance of RPRI radars.

For RPRI radar, a received signal is non-stationary and the traditional Doppler spectrum spread compensation methods [7-9] cannot work well. In addition, the traditional compensation methods are performed along azimuth direction and cannot resolve the RCM problem [10-12]. A new type of methods, such as Radon Fourier Transform (RFT) [5] and Radon-Fractional Fourier Transform (RFRFT) [13], have been proposed to compensate the RCM and Doppler spectrum spread during the integration time simultaneously. However, these methods cannot accumulate the energy completely due to the random jittering phase among different pulses. This problem can be treated as the non-uniform sampling and be solved by NUFRFT. The non-uniform sampling among different pulses can improve the anti-jamming ability of radar. In this paper, a new integration method based on Radon non-uniform Fractional Fourier Transform (Radon-NUFRFT) is proposed, which can correct the RCM with searching in the constructed motion parameter space and compensate the Doppler spectrum spread using NUFRFT as well. Simulation results demonstrate that the proposed method can obtain good detection performance even in low SNR scenario.

The remainder of this paper is organized as follows. Section II establishes the mathematical model of echo signal. Section III describes the proposed integration method based on Radon-NUFRFT. Section IV presents some simulated data to validate the proposed method. Section V concludes the paper.

## II. SIGNAL MODELING

Assume the radar adopts linear frequency modulated (LFM) waveforms,

$$s_T(t,\tau) = \text{rect}(\tau/T_p)\exp(j\pi\gamma\tau^2)\exp[j2\pi f_c(t+\tau)] \quad (1)$$

where rect(x) is the window function and equal to 1 for $|x| \le 1/2$, and 0, otherwise; $T_p$ is the pulse width; $f_c$ is the carrier frequency; $\gamma$ is the modulation rate; $\tau$ is the fast time, i.e., the range time; $t = [n + P(n')]T$, $(n = 0,1,\cdots N-1)$ is the slow time; $T$ is the average pulse repetition time; $N$ is the number of coherent integrated pulses; $P(n')$ is the random sequence within a certain range with $n'$ satisfying $n' = \mod(n,M)+1$; $M$ $(M < N)$ is the length of the random sequence, which has $M$ independent, identically distributed (i.i.d.) random samples uniformly distributed in $[-1,1]$. The random sequence is used to generate the PRI-jittering waveform in which the values of the PRIs have a slight jitter. This waveform is designed for anti-jamming. From the above description, it can be seen that the signal among different PRIs can be seen as periodic non-uniformly sampled signals with the period $MT$.

The received baseband signal after range compression can be expressed as

$$s(t,\tau) = \sigma G \text{sinc}\left[\pi B\left(\tau - 2R(t)/c\right)\right]\exp\left[-j4\pi R(t)/\lambda\right] \quad (2)$$

where $\sigma$ is the backscattering coefficient of a target, $G$ is the range compression gain, $B$ is the bandwidth of the signal, c is the speed of light, $\lambda = c/f_c$ is the wavelength, and

Tian Jing, Yang Gang, Cui Wei, and Wu Si-Liang are with the Radar Research Laboratory, Beijing Institute of Technology, Beijing 100081, China (e-mail: 10905046@bit.edu.cn).

Xia Xiang-Gen is with the Department of Electrical and Computer Engineering, University of Delaware, Newark, DE 19716 USA (e-mail: xxia@ee.udel.edu).

sinc$(x)=\sin(x)/x$.

The instantaneous slant range $R(t)$ between the platform and the target can be expressed as

$$R(t) = R_0 - v_0 t - a_0 t^2 /2 \quad (3)$$

Substituting (3) into (2) yields

$$\begin{aligned} s(t,\tau) &= \sigma G \text{sinc}\left[\pi B\left(\tau - 2\frac{R_0 - v_0 t - a_0 t^2/2}{c}\right)\right] \\ &\quad \cdot \exp\left[-j4\pi \frac{R_0 - v_0 t - a_0 t^2/2}{\lambda}\right] \\ &= \sigma G \text{sinc}\left[\pi B\left(\tau - 2\frac{R_0 - v_0[n+P(n')]T - a_0[(n+P(n'))T]^2/2}{c}\right)\right] \\ &\quad \cdot \exp\left[-j4\pi \frac{R_0}{\lambda}\right]\exp\left[j4\pi \frac{v_0 nT + a_0(nT)^2/2}{\lambda}\right] \\ &\quad \cdot \exp\left[j4\pi \frac{v_0 P(n')T + a_0\left[2nP(n')T^2 + (P(n')T)^2\right]/2}{\lambda}\right] \end{aligned}$$

(4)

In (4), it is obvious that the signal envelope varies with the slow time because of the target radial velocity and acceleration, and the RCM occurs when the offset of range exceeds the range resolution. Similarly, the Doppler spectrum spread occurs if the offset of frequency exceeds the frequency resolution due to the radial acceleration. In addition, it can also be seen from (4) that the phase of the azimuth signal fluctuates among different pulses, which can be decomposed into two parts, i.e., the fixed varying part $\varphi_1 = \exp\left\{j4\pi\left[v_0 nT + a_0(nT)^2/2\right]/\lambda\right\}$ and the jittering part $\varphi_2 = \exp\left\{j4\pi\left[v_0 P(n')T + a_0\left[2nP(n')T^2 + (P(n')T)^2\right]/2\right]/\lambda\right\}$.

The fixed phase variation among different pulses is caused by the variation of $nT$, while the jittering phase among different pulses is caused by the variation of PRI, i.e., $P(n')T$. In practice, since a moving target is non-cooperative, we cannot obtain its motion parameters and the problems of the RCM, Doppler spectrum spread and the phase fluctuations among different pulses of RPRI signals cannot concentrate the energy of the target completely. In the next section, we describe a new integration method, which can deal with those problems mentioned above, and thus realize the coherent integration in the RPRI radar.

### III. Radon-NUFRFT

In this section, the Radon-NUFRFT is proposed based on the FRFT for periodic non-uniformly sampled signals. Hence, we first briefly review the representation of FRFT for this type of signal. Then, the Radon-NUFRFT is proposed.

*A. FRFT of Non-uniformly Sampled Signals*

As a special kind of non-stationary signal, the chirp signal is widely used in signal processing areas like radar and sonar. FRFT is a useful tool for linear chirp signal processing. In many practical applications, non-uniform sampling occurs in many data acquisition systems due to imperfect timebase. If we still treat these sampled data as uniform samplings and apply the conventional operation on these data, the timebase errors cannot be ignored in the following processing [14]. Therefore, it is worthwhile exploring the fractional spectral representation of periodic non-uniformly sampled signals in the fractional Fourier domain.

Let $f(t)$ be a continuous time signal and $F_\alpha(u)$ be the FRFT of $f(t)$. $f(t_n)$ is obtained by sampling $f(t)$, which satisfies $f(t_n) = f(t)\sum_{n=-\infty}^{\infty} \delta(t-t_n)$. The FRFT of $f(t_n)$ is a sampled version of the FRFT for continuous time signal:

$$F_\alpha[f(t_n)](u) = \sum_{n=-\infty}^{\infty} K_\alpha(u,t_n) f(t_n) \quad (5)$$

where

$$K_\alpha(u,t_n) = \begin{cases} A_\alpha \exp\left\{jt_n^2 \cot\alpha/2 - jut_n \csc\alpha + ju^2 \cot\alpha/2\right\}, \alpha \neq k\pi \\ \delta(t_n - u), \quad \alpha = 2k\pi \\ \delta(t_n + u), \quad \alpha + \pi = 2k\pi \end{cases}$$

, $A_\alpha = \sqrt{(1-j\cot\alpha)/2\pi}$, $\alpha = p\pi/2$ is the fractional angle, and $p$ is the fractional order.

For the signal $f(t_n)$ with uniform sampling time intervals, i.e., $t_n = nT_s$, (5) can be equivalently represented as

$$F_\alpha[f(t_n)](u) = \frac{1}{T_s} e^{j\frac{u^2}{2}\cot\alpha} \cdot \left\{F_\alpha(u) e^{-j\frac{u^2}{2}\cot\alpha} * \sum_{n=-\infty}^{\infty} \delta\left[u - n\frac{2\pi\sin\alpha}{T_s}\right]\right\} \quad (6)$$

where $T_s$ denotes the sampling interval and $*$ denotes complex conjugate operator. (6) shows the relationship between the FRFT of continuous signal and the FRFT of its uniformly sampled signal. A fast computation of $F_\alpha[f(t_n)](u)$ can be found in [15-17].

Periodic non-uniformly samplings $f(t_n)$ satisfy $t_n = t_{km} = kMT + t_m, m = 0,1,2,...,M-1, k \in Z$, where $t_n$ is the periodic non-uniform sampling time with a recurrent period $MT$ and has $M$ non-uniform sampling points within each period, $t_m = mT + r_m T$ is the $m$ th sampling time within a period, $r_m T$ is the sampling time offset and $T$ is the average sampling period [14]. For the efficient calculation of the FRFT of $f(t_n)$, one type of method is to represent the FRFT of $f(t_n)$ in terms of the FRFT of uniformly sampled subsequences. The basic principle is to decompose the non-uniformly sampled sequence into several uniformly sampled subsequences and to obtain the FRFT representation of the original sequence through summing up the efficiently computed FRFT of the uniformly sampled subsequences. The detailed steps to calculate $F_\alpha[f(t_n)](u)$ is described in [18] and the main steps are as follows.

The sampling points $f(t_n)$ are divided into several groups of points. The $m$ th sampling group can be seen as uniformly sampling signal with sampling period $MT$, i.e.

$s_m = [f(t_m), f(t_{M+m}), f(t_{2M+m}), ...]$ and the total sampling sequence can be represented as $s = \{f(t_n) = f(t_{km}) | t_{km} = kMT + t_m, m = 0,1,2,...,M-1, k \in Z\}$. Then the original sequence can be further denoted as $s = \sum_{m=0}^{M-1} \bar{s}_m z^{-m}$, where $z^{-1}$ is the unit delay operator, $\bar{s}_m = [f(t_m), 0, ..., (M-1) \text{zeros}, f(t_{M+m}), 0, 0, ...], m = 0, 1, ..., M-1$ and

$\bar{s}_m z^{-m} = [(m \text{ zeros}), f(t_m), 0, ..., (M-1) \text{zeros}, f(t_{M+m}), 0, 0, ...]$

is obtained by shifting $\bar{s}_m$ $mT$ positions to the right. Therefore, the discrete fractional spectrum of the periodic non-uniformly sampled signal sequence $s$ can be derived by the summation of the FRFT of $M$ subsequence $\bar{s}_m z^{-m}$.

According to (6), the FRFT of $\bar{s}_m$ can be derived as

$$F_\alpha[\bar{s}_m](u) = \frac{1}{MT} e^{j\frac{u^2}{2}\cot\alpha} \cdot \left\{ F_\alpha[f(t+t_m)](u) e^{-j\frac{u^2}{2}\cot\alpha} * \sum_{n=-\infty}^{\infty} \delta\left[u - n\frac{2\pi\sin\alpha}{MT}\right] \right\} \quad (7)$$

Based on the property of FRFT, the expression of $F_\alpha[f(t+t_m)](u)$ can be derived as

$$F_\alpha[f(t+t_m)](u) = F_\alpha[f(t)](u+t_m\cos\alpha) e^{j\left[\frac{1}{2}(t_m^2 \sin\alpha\cos\alpha + ut_m\sin\alpha)\right]} \quad (8)$$

Based on (7) and (8), the FRFT of $s = \sum_{m=0}^{M-1} \bar{s}_m z^{-m}$ can be further derived as

$$F_\alpha[s](u) = \frac{1}{MT} \sum_{m=0}^{M-1} e^{j\frac{(u-mT\cos\alpha)^2}{2}\cot\alpha} e^{j\frac{1}{2}(mT)^2 \sin\alpha\cos\alpha - mTu\sin\alpha}$$
$$\times \left\{ \sum_{n=-\infty}^{\infty} F_\alpha\left(u - mT\cos\alpha - n\frac{2\pi\sin\alpha}{MT} + t_m\cos\alpha\right) \right. \quad (9)$$
$$\left. \times e^{j\left[\frac{1}{2}\left(t_m^2 \sin\alpha\cos\alpha + \left(u-mT\cos\alpha - n\frac{2\pi\sin\alpha}{MT}\right)t_m\sin\alpha\right)\right]} e^{-j\frac{\cot\alpha}{2}\left(u-mT\cos\alpha-n\frac{2\pi\sin\alpha}{MT}\right)^2} \right\}$$

Since $r_m = (mT - t_m)/T$, (9) can be simplified as [18]

$$F_\alpha[s](u) = \frac{1}{MT} \sum_{n=-\infty}^{\infty} \sum_{m=0}^{M-1} F_\alpha\left[u - n\frac{2\pi\sin\alpha}{MT} - r_m T\cos\alpha\right] \quad (10)$$
$$\times e^{j\frac{1}{2}T^2 r_m^2 \sin\alpha\cos\alpha} e^{j\frac{2\pi n}{MT}\cos\alpha\left(u-\frac{\pi n\sin\alpha}{MT}\right)} e^{-jr_m T\sin\alpha\left(u-\frac{2\pi n\sin\alpha}{MT}\right)} e^{-j\frac{2\pi nm}{M}}$$

Then the FRFT of periodic non-uniformly samplings $f(t_n)$ can be calculated by (10).

To distinguish from the FRFT of the uniform samplings, the FRFT of the periodic non-uniformly samplings of $f(t)$ is denoted as NUFRFT in this paper for simplicity. And the NUFRFT of $f(t)$ is defined as

$$F_{NUFRFT}(\alpha, u) = \text{NUFRFT}[f(t)]$$
$$= \frac{1}{MT} \sum_{n=-\infty}^{\infty} \sum_{m=0}^{M-1} F_\alpha\left[u - n\frac{2\pi\sin\alpha}{MT} - r_m T\cos\alpha\right] \quad (11)$$
$$\times e^{j\frac{1}{2}T^2 r_m^2 \sin\alpha\cos\alpha} e^{j\frac{2\pi n}{MT}\cos\alpha\left(u-\frac{\pi n\sin\alpha}{MT}\right)} e^{-jr_m T\sin\alpha\left(u-\frac{2\pi n\sin\alpha}{MT}\right)} e^{-j\frac{2\pi nm}{M}}$$

where NUFRFT$(\cdot)$ is the NUFRFT operator on a function of $t$.

The FRFT of a chirp signal $f(t) = D\exp\left[j2\pi(f_0 t + m_0 t^2/2)\right]$ with angle $\alpha$ can be derived as

$$F_\alpha(u) = 2\pi D A_\alpha e^{j\frac{u^2}{2}\cot\alpha} \delta(f_0 - u\csc\alpha) \quad (12)$$

where $D$ is a constant, $A_\alpha = \sqrt{(1-j\cot\alpha)/2\pi}$ and $\cot\alpha + m_0 = 0$.

Substituting (12) into (11), the NUFRFT of the chirp signal $f(t)$ can be represented as

$$F_{NUFRFT}(\alpha, u) = \text{NUFRFT}[f(t)]$$
$$= \frac{2\pi D A_\alpha}{T} e^{j\frac{u^2}{2}\cot\alpha} \sum_{k=-\infty}^{\infty} \delta\left[f_0 - u\csc\alpha + k\frac{2\pi}{MT}\right] A(k) \quad (13)$$

where $A(k) = \frac{1}{M} \sum_{m=0}^{M-1} \left[ e^{-jr_m T\left(f_0 + \frac{1}{2}\cot\alpha T r_m\right) - jMT r_m T\cot\alpha} \right] e^{-jmk\frac{2\pi}{M}}$.

### B. Definition of Radon-NUFRFT

Based on the analysis above, we propose a novel transform, namely the Radon-NUFRFT, to achieve the coherent integration for the RPRI radar. Without loss of generality, the definition of Radon-NUFRFT is given as follows.

Suppose that $s(t,r) \in \mathbb{C}$ is a two dimensional complex function defined in the $(t,r)$ plane and the line equation $r(t) = r - vt - at^2/2$ represents motion with acceleration, which is used for searching lines in the plane. Then the Radon-NUFRFT is defined as

$$S_{RNUFRFT}(\alpha, u; r, v, a) = \text{NUFRFT}[s(t, 2r(t)/c)]$$
$$= \text{NUFRFT}\left[s\left(t, 2(r - vt - at^2/2)/c\right)\right] \quad (14)$$

where $(r, v, a)$ denote the searching parameters.

Specifically, for the signal shown in (4), when the searching trajectory coincides with the target trajectory, i.e., $r(t) = R_0 - v_0 t - a_0 t^2/2$ is satisfied for each PRI, the peak of its envelope is reached and its Radon-NUFRFT is given as

$$S_{RNUFRFT}(\alpha, u; R_0, v_0, a_0) = \text{NUFRFT}[s(t, 2r(t)/c)]$$
$$= \text{NUFRFT}\left[\begin{array}{l} \sigma G \text{sinc}\left[\pi B\left(\frac{2r(t)}{c} - 2\frac{R_0 - v_0 t - a_0 t^2/2}{c}\right)\right] \\ \cdot \exp\left[-j4\pi\frac{R_0 - v_0 t - a_0 t^2/2}{\lambda}\right] \end{array}\right] \quad (15)$$
$$= \text{NUFRFT}\left[\sigma G \exp\left(-j4\pi\frac{R_0}{\lambda}\right) \exp\left(j4\pi\frac{v_0 t + a_0 t^2/2}{\lambda}\right)\right]$$

Denoting $E$ by $E = \sigma G \exp(-j4\pi R_0/\lambda)$, then according to (11) and (13), we can further obtain

$$S_{RNUFRFT}(\alpha, u; R_0, v_0, a_0) = \frac{2\pi E A_\alpha}{T} e^{j\frac{1}{2}\cot\alpha u^2}$$
$$\cdot \sum_{k=-\infty}^{\infty} \delta\left[\frac{2v_0}{\lambda} - u\csc\alpha + k\frac{2\pi}{MT}\right] A(k) \quad (16)$$

$$A(k) = \frac{1}{M}\sum_{m=0}^{M-1}\left[e^{-jr_m T\left(\frac{2v_0}{\lambda}+\frac{1}{2}\cot\alpha T r_m\right)-jMT r_m T \cot\alpha}\right]e^{-jmk\frac{2\pi}{M}}$$

and

$$A_\alpha = \sqrt{(1-j\cot\alpha)/2\pi}.$$

According to the above analysis, only when the searching initial range, searching radial velocity and searching radial acceleration are, respectively, equal to the real range, radial velocity and radial acceleration of the target, the energy can be accumulated completely in the FRFT domain by searching a proper rotation angle and the peak value of $S_{RNUFRFT}(\alpha, u; R_0, v_0, a_0)$ can be obtained through peak searching. Then the parameter estimates of the target can be obtained by

$$\begin{cases} [\hat{\alpha}_0, \hat{u}_0] = \arg\max_{\alpha,u}|S_{RNUFRFT}(\alpha,u;R_0,v_0,a_0)| \\ \hat{a}_0 = -\lambda \cot(\hat{\alpha}_0)/(2S^2) \\ \hat{v}_0 = \lambda \hat{u}_0 \csc(\hat{\alpha}_0)/(2S) \end{cases} \quad (17)$$

where $S = \sqrt{T_{total}/\bar{f}_r}$ is the normalized scaling factor for dimensional normalization [19], $T_{total}$ is the total integration time and $\bar{f}_r$ is the average pulse repetition frequency.

### C. Comparisons among Radon-NUFRFT, MTD, RFT, and RFRFT

Before comparing the proposed Radon-NUFRFT with MTD, RFT, and RFRFT, the definitions of MTD and RFT are introduced briefly.

The MTD process is shown as follows

$$S_{MTD}(f,\tau) = \sum_{n=0}^{N-1} s(t_n,\tau)\exp(-j2\pi f t_n) \quad (18)$$

where $s(t_n,\tau)$ is the signal after range compression and $t_n = nT$. For the target with uniform linear motion, the above coherent integration in one range cell is the well-known MTD method, which is also regarded as a Doppler filter bank. The Doppler shift can be determined by the maximum value of different filters. Unfortunately, $s(t,\tau)$ shown in (4) has the problem of RCM, the Doppler spectrum spread and the phase fluctuation of the azimuth signal among different pulses, the FFT-based MTD method becomes invalid.

RFT is proposed to extract the observation values in the two-dimensional range versus slow-time plane according to the motion parameters and finally accumulate the target's energy as a peak by integrating these observations with discrete Fourier Transform. The RFT process can be represented as [5]

$$S_{RFT}(r,v,a) = \sum_{n=0}^{N-1} s\left(t_n, 2(r-vt_n-at_n^2/2)/c\right) \cdot \exp\left[-j4\pi(vt_n+at_n^2/2)/\lambda\right] \quad (19)$$

where $t_n = nT$, $r_s(t_n) = r - vt_n - at_n^2/2$ is the moving trajectory of the target to be searched for and $(r,v,a)$ denote the searching parameters. The RFT can resolve the RCM and the Doppler spectrum spread problem and generate the ultimate coherent peak when the searching parameters $(r,v,a)$ are equal to the real parameters $(R_0,v_0,a_0)$ of a target. For the signal shown in (4), since the slow time is $t = t_n + P(n')T$, the phase $\varphi_1 = \exp\left[j4\pi(v_0 t_n + a_0 t_n^2/2)/\lambda\right]$ in (4) can be compensated by the compensation function $\exp\left[-j4\pi(vt_n+at_n^2/2)/\lambda\right]$ as shown in (19) when the searching parameters $(r,v,a)$ are equal to the real parameters $(R_0,v_0,a_0)$ of a target. However, the jittering phase $\varphi_2 = \exp\left\{j4\pi\left[v_0 P(n')T + a_0\left[2nP(n')T^2 + (P(n')T)^2\right]/2\right]/\lambda\right\}$ cannot be compensated since its time variable is jittering among different pulses, which is unmatched with the time of the constructed compensation function $\exp\left[-j4\pi(vt_n+at_n^2/2)/\lambda\right]$, and thus makes the target phase misaligned and the RFT method becomes invalid.

The Radon-NUFRFT of (4) can be expressed as shown in (20), at the bottom of the page, where $(r,v,a)$ denote the searching parameters and the kernel function

$$K_\alpha(u, t_n + P(n')T) = A_\alpha \exp(j\pi u^2 \cot\alpha) \\ \cdot \exp\left[j\pi(t_n+P(n')T)^2 \cot\alpha - j2\pi u(t_n+P(n')T)\csc\alpha\right] \quad (20)$$

is actually a chirp with the chirp rate $\cot\alpha$ and the initial frequency $-u\csc\alpha$ in terms of variable $t_n + P(n')T$. From (20), one can see that when $\alpha = -\operatorname{arccot}(2a_0/\lambda) + \pi$ and $u = -2v_0 \sin\alpha/\lambda$, the signal $\exp\left\{-j4\pi\left[R_0 - v_0(t_n+P(n')T) - a_0(t_n+P(n')T)^2/2\right]/\lambda\right\}$ is matched by $K_\alpha(u, t_n+P(n')T)$ in (20). And when the searching parameters $(r,v,a)$ are equal to the real parameters $(R_0,v_0,a_0)$ of a target, (20) becomes

$$S_{RNUFRFT}(\alpha,u;r,v,a) \\ = \mathrm{NUFRFT}\left\{s\left[(t_n+P(n')T), 2(r-v(t_n+P(n')T)-a(t_n+P(n')T)^2/2)/c\right]\right\} \\ = \sum_{n=0}^{N-1}\sigma G \mathrm{sinc}\left[\pi B\left(2\frac{r-v(t_n+P(n')T)-a(t_n+P(n')T)^2/2}{c} - 2\frac{R_0-v_0(t_n+P(n')T)-a_0(t_n+P(n')T)^2/2}{c}\right)\right] \\ \cdot \exp\left[-j4\pi\frac{R_0-v_0(t_n+P(n')T)-a_0(t_n+P(n')T)^2/2}{\lambda}\right] K_\alpha(u,t_n+P(n')T) \quad (21)$$

$$S_{RNUFRFT}(\alpha,u;R_0,v_0,a_0)$$

$$= \sum_{n=0}^{N-1} \sigma G \exp\left[-j4\pi \frac{R_0 - v_0(t_n + P(n')T) - a_0(t_n + P(n')T)^2/2}{\lambda}\right]$$
$$\cdot K_\alpha(u, t_n + P(n')T) \quad (22)$$

$$= \sum_{n=0}^{N-1} P \exp\left[j4\pi \frac{v_0(t_n + P(n')T) + a_0(t_n + P(n')T)^2/2}{\lambda}\right]$$
$$\cdot K_\alpha(u, t_n + P(n')T)$$

where $P = \sigma G \exp(-j4\pi R_0/\lambda)$.

Thus, from (16), when $\alpha = -\text{arccot}(2a_0/\lambda) + \pi$ and $u = -2v_0 \sin\alpha/\lambda$, (22) becomes

$$S_{RNUFRFT}(\alpha,u;R_0,v_0,a_0) = \frac{2\pi P A_\alpha}{T} e^{j\frac{1}{2}\cot\alpha u^2}$$
$$\cdot \sum_{k=-\infty}^{\infty} \delta\left[\frac{2v_0}{\lambda} - u\csc\alpha + k\frac{2\pi}{MT}\right] A(k) \quad (23)$$

where $A(k) = \frac{1}{M}\sum_{m=0}^{M-1} \left[e^{-jr_m T\left(\frac{2v_0}{\lambda} + \frac{1}{2}\cot\alpha Tr_m\right) - jMTr_m T\cot\alpha}\right] e^{-jmk\frac{2\pi}{M}}$ is the complex amplitude and $A_\alpha = \sqrt{(1-j\cot\alpha)/2\pi}$.

In addition, when the chirp rate and the initial frequency of a certain chirp basis with proper $\alpha$ are matched with the chirp rate and the initial frequency corresponding to the target motion parameters, respectively, the phase fluctuations among different pulses can be compensated and the target's energy can be accumulated at this chirp basis, which is represented as the delta function in (23). The acceleration and velocity of target can be obtained according to the coordinate of the peak in this NUFRFT domain.

For comparison, the RFRFT process can be expressed as [13]

$$S_{RFRFT}(\alpha,u;R_0,v_0,a_0) = \text{FRFT}\left[s\left(t_n, 2(R_0 - v_0 t_n - a_0 t_n^2/2)/c\right)\right]$$
$$= \sum_{n=0}^{N-1} s\left(t_n, 2(R_0 - v_0 t_n - a_0 t_n^2/2)/c\right) K_\alpha(u, t_n) \quad (24)$$

where $K_\alpha(u,t_n)$ is the kernel function of FRFT, which satisfies $K_\alpha(u,t_n) = A_\alpha \exp(j\pi u^2 \cot\alpha)\exp(j\pi t_n^2 \cot\alpha - j2\pi u t_n \csc\alpha)$ and $t_n = nT$. For the signal shown in (4), the phase $\varphi_1 = \exp\left[j4\pi(v_0 t_n + a_0 t_n^2/2)/\lambda\right]$ can be compensated in a proper rotation angle when the target parameters $(2v_0/\lambda, 2a_0/\lambda)$ are matched with a chirp basis $(-u\csc\alpha, \cot\alpha)$ of $K_\alpha(u,t_n)$. However, the jittering phase $\varphi_2 = \exp\left\{j4\pi\left[v_0 P(n')T + a_0\left[2nP(n')T^2 + (P(n')T)^2\right]/2\right]/\lambda\right\}$ cannot be matched with any a chirp basis of $K_\alpha(u,t_n)$ completely since its time variable is jittering among different pulses, thus makes the target phase misaligned and degrades the performance of RFRFT.

Compared with MTD, RFT and RFRFT, the advantages and differences of Radon-NUFRFT are summarized as follows:

1) The kernel of Radon-NUFRFT is introduced to act as an NUFRFT integration component. The azimuth signal along the searching trajectory is treated as a chirp signal with the non-uniform samplings, the parameters of which can be estimated by NUFRFT, as shown in (17).

2) MTD, RFT, and RFRFT cannot realize coherent integration for RPRI signals because of the random jittering phase among different pulses, while Radon-NUFRFT can perform well and acquire high anti-noise performance since it can realize the coherent integration among different pulses and thus generate the ultimate coherent peak according to (16) and (17). This will also be demonstrated in Section IV. Radon-NUFRFT not only has the same integration time as RFT and RFRFT (all longer than MTD), but also works well for the moving targets with RCM and Doppler spectrum spread, thereby improves the integration gain and detection performance.

3) The Radon-NUFRFT can be regarded as a special Doppler filter bank composed of filters with different fractional angles, which can simultaneously compensate and represent the velocity and acceleration.

4) Similarly to MTD, RFT, and RFRFT, Radon-NUFRFT can also be used to achieve the coherent integration for multiple targets. Furthermore, if the scattering intensities of different targets differ significantly, the CLEAN technique [20] can be employed to eliminate the effect of a strong target. In this way, the coherent integration of strong and weak moving targets can be achieved iteratively.

5) Compared with MTD, RFT, and RFRFT, Radon-NUFRFT has better performance but requires more computational complexity for coherent integration of moving targets. Specifically, the increased complexity results from several factors: (i) The integration time of Radon-NUFRFT is much longer than MTD processing, which indicates a larger number of samples involved in computations; (ii) Since the NUFRFT is obtained by summing up the FRFT of several subsequences, the computational burden is further increased; (iii) The Radon-NUFRFT is also time-consuming due to the searching operation with multiple fractional angles. The detailed computational complexity will be demonstrated in the following section.

D. *Procedure of the Coherent Integration Algorithm based on Radon-NUFRFT*

Based on the above analysis, the proposed Radon-NUFRFT is composed of the following five steps.

Step 1: Apply pulse compression on the received signal to accumulate energy within each pulse.

Step 2: According to the searching range and intervals of initial range, velocity and acceleration, the searching trajectory is determined by

$$r(t) = r_i - v_p t - a_q t^2/2 \quad (25)$$

where $r_i \in [r_{\min}, r_{\max}]$, $i = 1, 2, ..., N_r$, $v_p \in [-v_{\max}, v_{\max}]$, $p = 1, 2, ..., N_v$, $a_q \in [-a_{\max}, a_{\max}]$, $q = 1, 2, ..., N_a$, $N_r$, $N_v$ and $N_a$ denote the numbers of searches of range, velocity and acceleration, respectively.

Step 3: Perform Radon-NUFRFT on the searching trajectory

to realize the coherent integration.

Step 4: Repeat Step 3 for all the searching trajectories and obtain the integration outputs in the Radon-NUFRFT domain.

Step 5: Make a decision and obtain parameter estimates by searching for the peaks of the Radon-NUFRFT.

## IV. NUMERICAL EXAMPLES

In this section, some results with simulated data are presented to validate the proposed Radon-NUFRFT algorithm and performance comparison is also performed between the proposed Radon-NUFRFT and MTD, RFT, RFRFT. The simulated parameters and the target parameters are listed in TABLE 1 and TABLE 2.

TABLE 1 SYSTEM PARAMETERS OF RADAR

| System parameters (Unit) | Values |
|---|---|
| Carrier frequency (GHz) | 2.5 |
| Pulse width (us) | 10 |
| Bandwidth (MHz) | 20 |
| Sampling frequency (MHz) | 50 |
| Average pulse repetition interval (us) | 500 |
| Coherent integrated pulses | 1024 |

TABLE 2 INITIAL POSITIONS AND MOTION PARAMETERS OF TWO TARGETS

|  | Initial range (km) | Radial velocity (m/s) | Radial acceleration (m/s$^2$) |
|---|---|---|---|
| Target 1 | 50 | 51 | 9 |
| Target 2 | 50.15 | 45 | 12 |

### A. Coherent Integration for A Single Target

The signal is embedded in complex white Gaussian noise and the input SNR of target 1 is SNR = −23 dB. Fig. 1(a) shows the result of the signal after pulse processing, which shows that the target is buried in the noise. Fig. 1(b) shows the result of the signal without noise after pulse compression processing. It is obvious that the signal energy of the target spreads over several range cells. Fig. 1(c) illustrates the fractional spectrum of Radon-NUFRFT in terms of the searching fractional order. It can be seen that the energy of the target can be accumulated completely and based on the peak detection, we can obtain that the optimal searching fractional order is 1.024 and the corresponding frequency is 13.59. Then we can calculate the estimates of the targets with the values of $R_{T1} = 50$ km, $v_{T1} = 51$ m/s, $a_{T1} = 9$ m/s$^2$ according to (17).

### B. Coherent Integration for Multiple Targets

In this subsection, the coherent integration performance is evaluated for MTD, RFT [5], RFRFT and Radon-NUFRFT. The parameters used in the simulation are listed in Table 1 and Table 2. The signal is embedded in complex white Gaussian noise and the input SNR of the two targets are $SNR_1 = -20$ dB and $SNR_2 = -25$ dB. Fig. 2(a) shows the result after pulse compression in the $t-r$ domain, in which one target trajectory is clear while the other one is blurry. Fig. 2(b) shows the result of the signal without noise after pulse compression processing. It is obvious that the RCM occurs for both targets. Fig. 2(c) describes the result of MTD, which shows that the target energy cannot be accumulated completely because of the RCM and Doppler spectrum spread. Figs. 2(d), 2(e) and 2(f) show the results of RFT, RFRFT and Radon-NUFRFT, respectively. It can be seen that Radon-NUFRFT has larger output amplitude than RFT and RFRFT.

### C. Detection Performance

1000 trials are performed to evaluate the detection performance of target 1 for MTD, RFT, RFRFT, and Radon-NUFRFT. The signal is embedded in complex white Gaussian noise and the constant false alarm (CFAR) detector is applied for the four methods. The false alarm ratio is set to be $P_{fa} = 10^{-2}$. Fig.3 shows the detection probability versus the input SNR for the four methods. As shown in Fig.3, Radon-NUFRFT performs well even in low SNR scenarios and the detection performance of Radon-NUFRFT is superior to MTD, RFT, and RFRFT thanks to its ability to deal with the RCM, Doppler spectrum spread and the phase fluctuations among different pulses as well as the better performance on signal concentration.

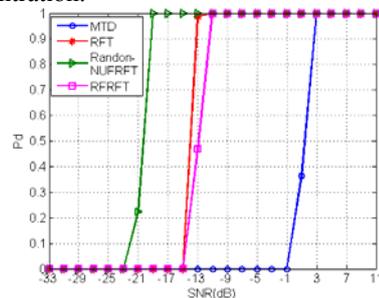

Fig.3. Detection probability versus the input SNR for MTD, RFT RFRFT, and Radon-NUFRFT.

Finally, the computational complexities of the four methods are given. Under the same condition, the computing time of the four methods for one trial is shown in TABLE 3. From this TABLE, it is obvious that the proposed method costs more time than the others due to the reasons analyzed in Section III-C. The main configuration of the computer is as follows: CPU: Intel Core i7-3770S 3.1GHz; RAM: 10G; Operating System: Windows 7; Software: Matlab 2012b.

TABLE 3 COMPUTING TIME FOR THE FOUR METHODS

|  | MTD | RFT | RFRFT | Radon-NUFRFT |
|---|---|---|---|---|
| $t(s)^*$ | 0.006 | 24.117 | 33.453 | 49.918 |

## V. CONCLUSIONS

In this paper, we have introduced a coherent integration detection method, called Radon-NUFRFT, for RPRI radar. This method can compensate RCM and Doppler spectrum spread simultaneously over long integration time. It can realize data extraction for the signal after pulse compression through jointly searching along range, velocity and acceleration dimensions. Then NUFRFT is applied on the searching trajectory to realize coherent integration. Because of coherent integration, this method can estimate motion parameters with high accuracy even in low SNR scenarios, as shown by the simulation results. The detection performance of the proposed algorithm has been validated by experimental results, which shows that it has better detection performance than MTD, RFT and RFRFT.

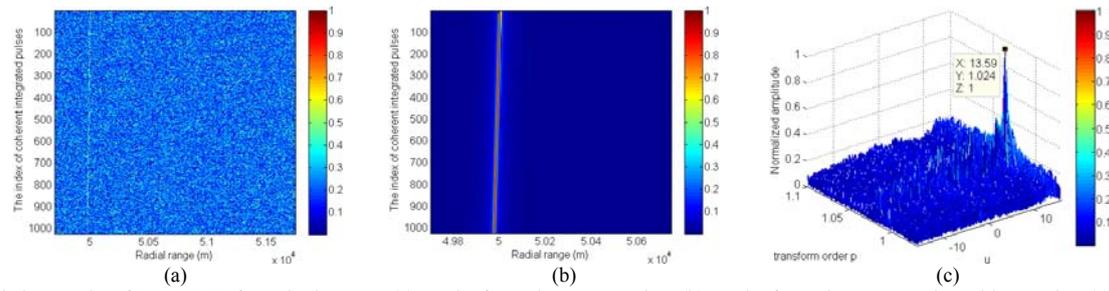

Fig.1. Simulation results of RNUFRFT for a single target. (a) result after pulse compression. (b) result after pulse compression without noise. (c) result of Radon-NUFRFT.

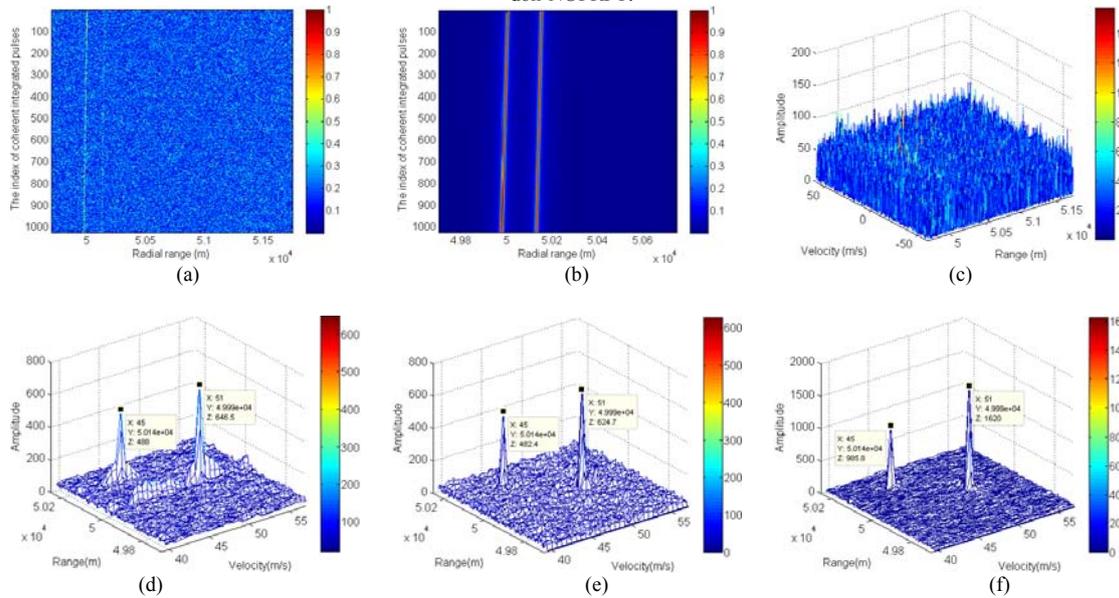

Fig.2. Simulation results via four methods for two targets. (a) result after pulse compression. (b) result after pulse compression without noise. (c) result of MTD. (d) result of RFT. (e) result of RFRFT. (e) result of Radon-NUFRFT.

## REFERENCES

[1] M. Kaveh, and G. R. Cooper, "Average ambiguity function for a randomly staggered pulse sequence," *IEEE Trans. Aerosp. Electron. Syst.,* vol. AES-12, no. 3, pp. 410-413, May 1976.

[2] R. Benjamin, "Form of Doppler processing for radars of random p.r.i. and r.f.," *Electron. Lett.*, vol. 15, no. 24, pp. 782, Nov. 1979.

[3] L. Vergara-Dominguez, "Analysis of the digital MTI filter with random PRI," *IEE Proc.-F*, vol. 140, no. 2, pp. 129-137, Apr. 1993.

[4] S. P. Sira, L. Ying, and A. Papandreou-Suppappola, "Waveform-agile sensing for tracking," *IEEE Signal Process. Mag.*, vol. 1, no. 26, pp. 53-64, Jan. 2009.

[5] Xu J, Yu J, Peng Y N, and X.-G. Xia, "Radon-Fourier transform for radar target detection (I): generalized Doppler filter bank," *IEEE Trans. Aerosp. Electron. Syst.,* vol. 47, no. 2, pp. 1186-1202, Apr. 2011.

[6] R. Tao, N. Zhang, and Y. Wang, "Analysing and compensating the effects of range and Doppler frequency migrations in linear frequency modulation pulse compression radar," *IET Radar Sonar Navig.,* vol. 5, no. 1, pp. 12-22, Jan. 2011.

[7] E. J. Kelly, "The radar measurement of range, velocity and acceleration," *IRE Trans. Military Electron.*, vol. 5, no. 2, pp. 51-57, Apr. 1961.

[8] P. J. Loughlin, J. W. Pitton, and L. E. Atlas, "Bilinear time-frequency representations: new insights and properties," *IEEE Trans. Signal Process.,* vol. 41, no. 2, pp. 750-767, Feb. 1993.

[9] S. Peleg, and B. Friedlander, "The discrete polynomial-phase transform," *IEEE Trans. Signal Process.,* vol. 43, no. 8, pp. 1901-1914, Aug. 1995.

[10] C. C. Chen, and H. C. Andrews, "Target-motion-induced radar imaging," *IEEE Trans. Aerosp. Electron. Syst.,* vol. 16, no. 1, pp. 2-14, Jan. 1980.

[11] R. P. Perry, R. C. DiPietro, and R. L. Fante, "SAR imaging of moving targets," *IEEE Trans. Aerosp. Electron. Syst.,* vol. 35, no. 1, pp. 188-200, Jan. 1999.

[12] F. Berizzi, and G. Corsini, "Autofocusing of inverse synthetic aperture radar images using contrast optimization," *IEEE Trans. Aerosp. Electron. Syst.,* vol. 32, no. 3, pp. 1185-1191, Jul. 1996.

[13] X. L. Chen, J. Guan, N. B. Liu, and Y. He, "Maneuvering target detection via Radon-fractional Fourier transform-based long-time coherent integration," *IEEE Trans. Signal Process.*, vol. 62, no. 4, pp. 939–953, Feb. 2014.

[14] Y.-C. Jenq, "Digital spectra of nonuniformly sampled signals: fundamentals and high-speed waveform digitizers," *IEEE Trans. Instrum. Meas.,* vol. 37, no. 2, pp.245-251, Jun. 1988.

[15] H. M. Ozaktas, O. Arýkan, A. Kutay, and G. Bozdadý, "Digital computation of the fractional Fourier transform," *IEEE Trans. Signal Process.*, vol. 44, no. 9, pp. 2141-2150, Sep. 1996.

[16] S.-C. Pei, and J.-J. Ding, "Closed-form discrete fractional and affine Fourier transform," *IEEE Trans. Signal Process.*, vol. 48, no. 5, pp. 1338-1353, May 2000.

[17] X. G. Deng, X. P. Li, D. Y. Fan, and Y. Qiu, "A fast algorithm for fractional Fourier transform," *Opt. Commun.*, vol. 138, no. 4, pp. 270-274, Jun. 1997.

[18] R. Tao, B. Z. Li, and Y. Wang, "Spectral analysis and reconstruction for periodic non-uniformly sampled signals in fractional Fourier domain," *IEEE Trans. Signal Process.,* vol. 55, no. 7, pp.3541-3547, Jul. 2007.

[19] H. M. Ozaktas, and M. A. Kutay, "Digital computation of the fractional Fourier transform," *IEEE Trans. Signal Process.,* vol. 44, no. 9, pp. 2141-2150, Jul. 1996.

[20] T. Jenho, and B. D. Steinberg, "Reduction of sidelobe and speckle artifacts in microwave imaging: the CLEAN technique," *IEEE Trans. Antennas Propagat.,* vol. 36, no. 4, pp. 543-556, Apr. 1988.